\documentclass[11pt,twoside]{article}
\usepackage{./asp2014}

\aspSuppressVolSlug
\resetcounters

\begin{document}

\title{Shellspec39 - a tool for modelling the spectra, light curves,
and images of interacting binaries and exoplanets}
\author{Budaj~J\'{a}n,$^1$
\affil{$^1$Astronomical Institute, Slovak Academy of Sciences, 
Tatransk\'{a} Lomnica, Slovak Republic; \email{budaj@ta3.sk}}}

% This section is for ADS Processing.  There must be one line per author.
\paperauthor{Budaj Jan}{budaj@ta3.sk}{}{Astronomical Institute}
{Slovak Academy of Sciences}{Tatranska Lomnica}{}
{05960}{Slovak Republic}
%\paperauthor{Sample~Author2}{Author2Email@email.edu}{ORCID_Or_Blank}{Author2 Institution}{Author2 Department}{City}{State/Province}{Postal Code}{Country}
%\paperauthor{Sample~Author3}{Author3Email@email.edu}{ORCID_Or_Blank}{Author3 Institution}{Author3 Department}{City}{State/Province}{Postal Code}{Country}

\begin{abstract}
Program SHELLSPEC is designed to calculate light curves, spectra, and 
images of interacting binaries and extra-solar planets immersed in 
a moving gaseous or dusty circumstellar matter. It solves simple 
radiative transfer along the line of sight in 3D moving media.
The Roche model and synthetic spectra from the stellar atmosphere models
such as TLUSTY from Ivan Hubeny can be used as a boundary condition for
the radiative transfer.
The latest publicly available version is Shellspec39.
The code has been combined with other methods such as Doppler tomography
and interferometry and used to analyze spectroscopic, photometric, and 
interferometric observations of binary stars and transiting exoplanets. 
A few examples are briefly mentioned.
\end{abstract}

\section{Introduction}

A significant fraction of stars are single and have the spherical shape.
The light we detect originates from their atmospheres.
There are sophisticated computer codes which calculate models
of the atmospheres such as TLUSTY \citep{hubeny88,hubeny95} and spectra 
emerging from such models such as SYNSPEC \citep{hubeny17}.
However, many stars are in binary systems and many of them form
close systems which triggers a mutual interaction between 
the components. As a result the shape of the objects departs from 
the sphere and acquires the Roche shape. 
Impinging irradiation heats the adjacent 
faces of the stars (planets) as well as is scattered or reflected
off the surfaces. Mass can be transferred between the components
or can escape from the system forming plethora of accretion structures
(streams, discs, jets, shells,...). These may give rise to various
emission and absorption lines with complicated profiles. 
These lines carry the information 
about the velocity field and state quantities such as temperatures and
densities.
To analyze such information 3D radiative transfer codes are being 
developed \citep{ibgui13}.
Such modelling is very complicated and demanding on the computer time 
and memory.
A certain simplification may be very useful, sometimes inevitable,
especially when one wants to solve the inverse problem of finding some
optimal properties of the model. That usually requires fitting of large 
volumes of observed data involving frequent on-fly recalculation
of the models.

With this in mind we developed a program called SHELLSPEC.
It might be used to calculate light curves, spectra, and 
images of interacting binaries and extra-solar planets immersed in 
a moving gaseous or dusty circumstellar matter (CM). It solves simple 
radiative transfer along the line of sight in 3D moving media.
The Roche model and synthetic spectra from the stellar atmosphere models
such as TLUSTY can be used as a boundary condition for
the radiative transfer.  The scattered light from the two stars can be 
taken into account assuming that CM is optically thin.
Scattering is non-isotropic and described by the proper phase function
which may be crucial especially in the case of Mie scattering on large 
dust grains.
The other assumptions include LTE and optional known state quantities 
and velocity fields in 3D.
These can be taken e.g. from the 3D hydrodynamic simulations.
Alternatively, a 3D model can be composed of a number of predefined
(non)transparent objects such as: a central star,
companion star, envelope, spot, stream, ring, disc, nebula, flow, jet,
ufo, or a shell which are described below.

At present, still a limited number of opacities in CM are taken into
account: line opacity with the natural, Stark, and Van der Waals 
broadening, an additional extra true absorption opacity
in the form of a table or an opacity distribution function as
a function of frequency and temperature,
HI bound-free and free-free opacity, H$^{-}$ bound-free and free-free,
Thomson scattering on free electrons, Rayleigh scattering on neutral
hydrogen, and Mie scattering and absorption on dust.
There is a 'quick and dirty' way how to include opacity of some
molecular species using those extra opacity distribution functions.
Corresponding thermal and scattering emissivities are included.
Model is defined in a 3D Cartesian grid and spectra are calculated
by ray tracing in another grid aligned with the observer's line of sight.
The code was written from scratch but we adopted a few subroutines
from other sources. Mainly partition functions
from the UCLSYN code \citep{smith88} and a few subroutines from 
the SYNSPEC code \citep{hubeny17}.
A brief description of the version No.39 follows below.

\section{Input}

Depending on the problem to be solved the user has to supply 
a proper input.
Here is the list of most important input files.
\begin{itemize}
\item 
shellspec.in - is the main input file where you define the 3D model.
It can be composed of a number of predefined structures briefly 
described in the next section or loaded from a separate file called 
shellspec.mod.
\item 
line.dat - atomic data for the spectral lines formed in the CM 
(Kurucz format, optional)
\item 
shellspec.mod - 3D model that contains for each point 
in space: temperature and density of the gas and dust, electron number
density is optional (it can be calculated in LTE), velocity vector,
and micro-turbulence. This can be e.g. an output of a 3D 
hydrodynamic simulation (optional).
\item 
abundances  - abundances of the CM (optional).
\item 
starspec1, starspec2, starspec3 - specify the grid of spectra of up to 
three nontransparent objects. This is the boundary condition for 
the radiative transfer in the CM. Default format is the output of
the model atmosphere codes TLUSTY and SYNSPEC (optional).
\item 
albedo1, albedo2 - albedo of the primary and secondary stars as 
a function of the wavelength.
Necessary for the reflection effect between the objects (optional).
\item 
dust\_opac, mie\_phase - these are tables with the dust opacities and
phase functions and are required only if there is a dust.
We pre-calculated such tables to be used within TLUSTY but they can be 
used also here \citep{budaj15} (optional).
\item 
gas\_opac - this is a table with molecular cross-sections as a function 
of wavelength. Default format is as in EXOMOL database
\citep{tennyson12} (optional).
\item 
chem\_eq\_tab - this is a table with molecular population for 
a particular molecule as a function of gas temperature, density, 
and chemical composition. It was calculated with the code {\AE}SOPUS by 
\cite{marigo09} (optional).
\end{itemize}

\section{Predefined structures}

As a simple alternative to a more sophisticated 3D model
there are a few simple predefined structures available which can be 
used to compose the 3D model. 
In case the structures happen to overlap in space a particular point
will bear the properties of the higher priority object.
Objects/structures ranked according to their priority are listed
and briefly described below.

%\begin{itemize}
%\item 
STAR:
a central nontransparent object which can rotate as a solid body
or possess a differential rotation with an optional inclination of 
the rotational axis and have a net space velocity. 
Can be treated as a black body or have its own spectrum. 
May be of the spherical or Roche shape.
Limb and gravity darkening can be applied to it. 
Spherical star may have a circular spot on the surface of different
temperature at a fixed location.
Irradiation effect on its surface (from the COMPANION)
can be considered.
The light scattered in the circumstellar medium which originates from 
this object can be taken into account
(neglecting its rotation, irradiation effect and assuming
spherical shape).  
Differential rotation applies only to the spherical shape and
does not affect the location of the spot. 
Designed to model mainly hotter or more luminous stellar components
as well as extra solar planets.

%\item
COMPANION: 
similar to the STAR.
Designed to model mainly a secondary (cooler or fainter component 
of a binary system).

%\item
ENVELOPE:
is an object enclosing the central STAR (or STAR and COMPANION). 
It is subject to the Roche shape and can be detached or contact. 
It rotates as a solid body, does not have limb nor gravity darkening, 
has constant temperature and densities.
It is possible to constrain this structure within a certain
height above and below the orbital plane. 
Designed to model envelopes and common envelopes of 
e.g. W UMa type stars.

%\item
SPOT: 
a spherical object which can rotate as a solid body with an optional 
inclination of the rotational axis and have a net space velocity. 
Designed to model a third body, spots on accretion discs, direct impact
regions, rotating circum-stellar shells. 

%\item
STREAM:
has the shape of a cylinder or cone with velocity varying linearly 
along the cone. 
Stream may rotate or be a subject of some rotational drag.
It may also have a net space velocity so that it can e.g. follow 
the movement of some associated star.
Density of the stream changes along the stream
to satisfy the continuity equation with some
modification to allow for modelling of additional phenomena.
Designed to model the mass transfer streams, outflows, or shadows.

%\item
RING:
is a circular ring or part of the ring (arc) with optional inclination 
and location. Mass in the center determines its Keplerian velocity. 
However, the velocity is uniform throughout the cross-section. 
The cross-section of the ring, $C$, has shape of 
a rectangle and may vary along the arc. Density, dust density and 
electron number density may change along the arc 
to satisfy the continuity equation and/or additional phenomena
e.g. dust destruction etc.
Designed to model rings around objects, arcs, comets.

%\item
DISC:
has either the shape of a rotating wedge 
(space complement to two opposite cones)
or of a slab, or of a rotational ellipsoid surrounding 
the central object with some mass.
It is farther constrained by two surfaces: its inner spherical
surface with radius $r_{in}$ and outer spherical or ellipsoidal surface
with radius $r_{out}$. This structure may have optional location
and inclination.
The velocity field of the disc depends on the mass of the central 
object, and is Keplerian within the disc plane.
Densities may vary in the radial direction as a power law.
Temperature may be either constant, or have a radial power law 
dependence, or obey a radial accretion disk structure of 
\cite{pringle81}.

%\item
NEBULA:
is another disk-like structure. It is located around the central object
with a certain mass,
has a Keplerian rotation, and may have also a net space velocity. 
It is defined in the cylindrical coordinates $(r,z)$.
Surface density, $\Sigma$, varies as a power law with the distance 
from the center.
Mid-plane density, $\rho_{0}$, is then determined as a function of 
distance from the surface density and vertical scale height, $H$.
Vertical scale height is a function of Keplerian velocity, $v$, and 
sound speed, $c_{s}$.
Vertical density behaviour is Gaussian with some modification
to allow for modelling of additional phenomena (e.g. wind).
Temperature structure in the radial direction is similar to DISC.
There is an option of a simple vertical gas temperature dependence
(e.g. inversion).
This is another option to model flared accretion or protoplanetary 
disks.

%\item
FLOW:
is identical to the STREAM but has a lower priority.
Designed to model the mass transfer streams, outflows,
or structures symmetric to the STREAM.

%\item
JET:
has the shape of one or two opposite cones emerging from the center.
It allows optional inclination and is farther limited by its inner, 
$r_{in}$, and outer, $r_{out}$, radii. It has constant temperature and
velocity along the beam. However, it may have a net space velocity so 
that it can move e.g. with the STAR. 
Densities vary along the jet to satisfy the continuity equation.
Designed to model mainly jets or, 
e.g., `shadows' cast by a cool, extended secondary from a more compact 
hot primary.

%\item
UFO:
is identical to the DISC but has lower priority. 
Intended to model an extension or atmosphere of 
the DISC or a second disc.

%\item
SHELL:
has the shape of a shell surrounding the central object. 
A few different velocity fields are built in.
The temperature is kept fixed and densities are either constant
or satisfy the continuity equation. It may have a net space velocity.
%\end{itemize}

\section{Output}

Here is a list of all output files.
\begin{itemize}
\item
shellspectrum - spectra of the model from different view points.
\item
lightcurve - light curve at different wavelengths obtained by combining
a sequence of spectra from different view points.
\item
fort.xx - are 2D images at some frequency from different view points.
\item
shellspec.out - a more detailed output of opacities, emissivities, 
optical depths, ... along some selected rays.
\end{itemize}

\section{Applications}

Possible application of this code include interacting binaries
\citep{budaj05,ghoreyshi11,gorlova12,bozic13,richards14}
and some exoplanets. 
For example,
\cite{tkachenko09,tkachenko10} converted an earlier version07
into Fortran90 an wrote an inverse program which was used to analyze
oscillating Algol type binaries.
\cite{chadima11} used it for post-processing of the 3D hydrodynamic 
simulations to study variability in the H{$\alpha$} emission
caused by a discontinuous mass transfer in binaries.
\cite{atwood12} combined the spectroscopic observations with 
the synthetic spectra and the Doppler tomography to analyze 
an Algol type binary AU Mon.
\cite{lehmann13} analyzed the properties of the eclipsing 
{$\delta$} Scuti star KIC 10661783.
\cite{bakis16} modelled the circumstellar material around the active 
binary R Arae.
\cite{sejnova16} studied a dynamical evolution of the disk of the Be 
star 60 Cygni.
\cite{garai18} used the code to analyze the Kepler light curve
of a peculiar extra-solar planet KOI 2700b featuring
variable asymmetric transits of its comet-like dusty tail.
Most recently Mirek Bro\v{z} and Janka Nemravov\'{a} wrote a package of 
codes in python that calculate interferometric observables from 
the Shellspec images and fit them simultaneously with the light curves.
The method was used to determine the properties of a bright binary
star $\beta$ Lyr \citep{mourard18}.
The reflection effect which is included in the Shellspec code 
\citep{budaj11} was now generalized and included in the PHOEBE 2
package for modelling eclipsing binaries and exoplanets 
\citep{horvat18}.

\section{Summary}

The latest publicly available version No.39 can be found here with 
the complete documentation, and example runs:\\
\url{http://www.ta3.sk/~budaj/shellspec}\\
A more detailed description is in \cite{budaj04} and in the updated
user manuals.
Dust phase functions and opacities can be found here with 
the description and references to the refractive index measurements 
adopted:\\
\url{http://www.ta3.sk/~budaj/dust}\\
Any comments, suggestions, or bug reports will be appreciated.
Thank you.

\acknowledgements The author would like to thank Ivan Hubeny
and late Mercedes Richards for their support and consultations over 
the years. The work was supported by the VEGA 2/0031/18,
APVV 15-0458, and Erasmus+ program (Per Aspera ad Astra Simul).

\bibliography{budaj}{}  % For BibTex

\end{document}